\documentclass[journal,twoside,web]{ieeecolor}
\usepackage{tmi}
\usepackage{cite}
\usepackage{amsmath,amssymb,amsfonts}
\usepackage{algorithmic}
\usepackage{graphicx}
\usepackage{textcomp}
\usepackage{stfloats}
\usepackage{threeparttable}
\usepackage{booktabs}
\usepackage{multirow}

\def\BibTeX{{\rm B\kern-.05em{\sc i\kern-.025em b}\kern-.08em
    T\kern-.1667em\lower.7ex\hbox{E}\kern-.125emX}}
\begin{document}
\title{Reconstruct high-resolution multi-focal plane images from a single 2D wide field image}
\author{Jiabo Ma, \IEEEmembership{Student, IEEE}, Sibo Liu, \IEEEmembership{Student, IEEE}, Shenghua Cheng, Xiuli Liu, Li Chen, Shaoqun Zeng
\thanks{Corresponding author: Shenghua Cheng.}
\thanks{J. Ma and S. Liu contributed equally to this work. }
\thanks{J. Ma, S. Liu, S. Cheng, X. Liu and S. Zeng are with Britton Chance Center for Biomedical Photonics, Wuhan National Laboratory for Optoelectronics, Huazhong University of Science and Technology, Wuhan, Hubei 430074, China and MoE Key Laboratory for Biomedical Photonics, School of Engineering Sciences, Huazhong University of Science and Technology, Wuhan, Hubei 430074, China. (email:
majiabo@hust.edu.cn; liam.liusibo@gmail.com; chengshen@hust.edu.cn; xlliu@mail.hust.edu.cn; sqzeng@mail.hust.edu.cn)}
\thanks{L. Chen is with Department of Clinical Laboratory, Tongji Hospital, Huazhong University of Science and Technology, Wuhan, Hubei 430030, China. (email: chenliisme@126.com)}}

\maketitle

\begin{abstract}
High-resolution 3D medical images are important for analysis and diagnosis, but axial scanning to acquire them is very time-consuming. In this paper, we propose a fast end-to-end multi-focal plane imaging network (MFPINet) to reconstruct high-resolution multi-focal plane images from a single 2D low-resolution wild filed image without relying on scanning. To acquire realistic MFP images fast, the proposed MFPINet adopts generative adversarial network framework and the strategies of post-sampling and refocusing all focal planes at one time. We conduct a series experiments on cytology microscopy images and demonstrate that MFPINet performs well on both axial refocusing and horizontal super resolution. Furthermore, MFPINet is approximately 24 times faster than current refocusing methods for reconstructing the same volume images. The proposed method has the potential to greatly increase the speed of high-resolution 3D imaging and expand the application of low-resolution wide-field images.

\end{abstract}


\section{Introduction}
\label{sec:introduction}

\IEEEPARstart{I}{n} the field of medical images, high-resolution (HR) 3D images are helpful for analysis and diagnosis \cite{sr_diag_0}. Therefore, how to obtain HR 3D images with high throughput is important. The main methods to obtain HR 3D optical images are to scan point by point or plane by plane through two-photon\cite{denk1990two}, confocal \cite{pawley2006handbook},  and optical sheet \cite{RN8}. By using these techniques, the information outside focal planes can be effectively suppressed from entering the imaging system, thus greatly improving the axial resolution of 3D optical images. But these methods are generally slow in imaging due to axial scanning. At the same time, phototoxicity and photobleaching will affect observed samples due to long time scanning \cite{phototoxity2017assessing}. In order to alleviate the above problem, some methods optimize scanning strategy\cite{RN6} and improve PSF\cite{RN11} to speed up imaging. With the development of computer technologies, 3D fluorescence information can also be obtained by non-scanning methods. These methods can map the axial information to 2D images at one time, such as fluorescence field microscope \cite{levoy2006light} and Fresnel hologram \cite{RN15}, then use iterative algorithms to reconstruct 3D images. These non-scanning methods are usually time-consuming in image reconstruction, and generally need specific optical elements or instruments as assistance, which increases the complexity of methods. 

In recent years, more and more methods apply deep learning to improve imaging resolution and increase imaging speed. Most super-resolution microscopy methods \cite{RN20, RN29, zhang2019high} implement on 2D images, which improves the resolution of images collected by low-end instruments and expands the application of low-resolution (LR) images in research and diagnosis. Further, some 3D super-resolution microscopy methods \cite{a-tensor,RN27,multi-contrast-mri,RN25} are also developed. 3D super-resolution technologies can greatly increase imaging speed while ensuring image resolution since the acquisition of 3D LR images requires fewer scans and produces less data. However, axial scanning is still required to acquire 3D images even if these super-resolution technologies are used. To overcome the problem, 2D-to-3D super-resolution methods are proposed, such as deep learning based virtual refocusing \cite{RN23}. The method can refocus a single wide-field image to acquire multi-focal plane (MFP) images. With the help of deep learning's powerful fitting ability and efficient forward inference, 3D images can be obtained fast without axial-scanning or special hardware.

However, the current method of obtaining MFP images from a single 2D wide-field image focuses on the axial refocusing and neglects the improvement of horizontal resolution. Therefore, the speed of acquiring HR MFP images still depends on horizontal fine scanning. To get rid of the dependence on HR images, in this paper, we propose a model called MFPINet (multi-focal plane imaging network) to reconstruct HR MFP wide-field images from a single LR 2D wide-field image based on deep learning. The designed MFPINet is end-to-end and does not need any iterative calculation. In order to generate realistic MFP images, the generative adversarial network (GAN) framework \cite{RN37} is introduced in the proposed MFPINet. Specially, the generator of MFPINet is modified from U-Net \cite{RN38} based on the characteristics of super resolution task. To reduce training difficulty and eliminate artifacts \cite{wang2018esrgan}, we introduce residual connections \cite{RN41} and remove common batch normalization (BN) \cite{RN40} layers. In addition, to obtain HR MFP images more efficiently, we further optimize the U-Net in two ways. First, for horizontal super resolution, we adopt a post-sampling strategy instead of up-sampling the input image, which greatly reduces memory consumption and computational complexity. Second, for axial refocusing, the designed generator outputs $N$ focal plane images at one time instead of outputting specific focal plane image each time and then stitching them together. 
As for discriminator, it is a common classification network used to distinguish real and fake MFP images. But to make it easier for discriminator to be aware of the difference of all focal planes, the entire MFP images are used as the input of discriminator. Since the input MFP images are 3D, to save memory, the axial information in the 3D image ($W\times H\times Z\times C$) is converted to the color space to obtain a multi-channel 2D image ($W\times H\times(Z\times C)$). Then 2D convolutions are used to process input images. In summary, we propose a fast end-to-end MFPINet to transform a LR 2D image into HR MFP images. No additional optical hardware is needed in the whole process, so it can be applied to all kinds of wide field images without major adjustment. We conduct a series of experiments on cytology images to demonstrate MFPINet can generate high-quality HR MFP images.

The main contributions of this paper are as follows: 
\begin{itemize}
	\item We first propose a generative model MFPINet for reconstructing high-resolution multi-focal plane images from a single 2D wide field image.
	\item We verify the effectiveness of MFPINet for both axial refocusing and horizontal super resolution on cytology images.
	\item Experimental results further show that MFPINet is approximately 24$\times$ faster than current refocusing methods.
\end{itemize}

\section{Related Works}
\subsection{2D-to-2D Super Resolution}
In recent years, the super resolution of 2D natural images \cite{dong2015image, lim2017enhanced, ledig2017photo, guo2020closed} has been widely studied. Dong et al. \cite{dong2015image} introduced convolutional neural network into super resolution, making deep learning play a key role in super resolution. Kim et al. proposed VDSR \cite{kim2016accurate}, which greatly improved the effect of super resolution due to using deeper network and a strategy of learning global residuals. EDSR \cite{lim2017enhanced} removed unnecessary modules of conventional residual networks, and achieved better results by expanding model architecture and stabilizing training process. To obtain more realistic HR images, Ledig et al. \cite{ledig2017photo} introduced GAN into super resolution and proposed SRGAN to obtain super-resolved images with richer details. 
But the introduction of GAN caused more artifacts. To alleviate artifacts, ESRGAN \cite{wang2018esrgan} introduced residual-in-residual dense blocks and removed batch normalization, which further improved the quality of reconstructed HR images.

At the same time, super-resolution methods designed for various medical images have also emerged. Wang et al. \cite{RN29} used an U-Net-like GAN model to improve the resolution of low numerical aperture(NA) fluorescence microscopy images. Ma et al. \cite{RN20} proposed a multi-stage and multi-supervised model PathSRGAN to expand the application of 4x/0.1-NA pathology images. Rivenson et al. \cite{rivenson2018deep} extended deep learning based super-resolution to mobile-phone-based microscopy images.

\subsection{3D-to-3D super resolution}
For 3D medical images, Pham et al. \cite{RN25} proved that LR 3D MRI images can be reconstructed into high-quality HR 3D MRI images by using deep convolution neural networks, which greatly improves the speed of obtaining HR MRI images. Zhou et al. \cite{zhou20203d} improved resolution of 3D fluorescence microscopic images by using a dual super resolution model. You et al. \cite{RN27} proposed a semi-supervised 3D CT super-resolution method by introducing Wassertein constraints to GAN and node constraints to loss functions. In summary, the above 3D super-resolution methods improves resolution and imaging speed of 3D medical images.

\subsection{2D-to-3D super resolution} 
Although 3D super resolution is helpful in expanding the application of LR medical images, the acquisition of 3D LR images still requires time-consuming axial scanning. With the progress of computational imaging, deep learning based 2D-to-3D super-resolution methods are gradually emerging. Page et al. \cite{page2020learning} proposed LFMNet to reconstruct confocal microscopy stacks from single light filed images. Wu et al. \cite{RN23} proposed Deep-Z model based on GAN, which can refocus single 2D wide-field fluorescence images to obtain 3D images. 

Although both MFPINet and Deep-Z can reconstruct 3D images from a single 2D wide field image, they are very different. First, MFPINet enhances the horizontal and axial resolution at the same time. So it does not require HR 2D images as input, which greatly improves the imaging speed and expands the applications of MFPINet. Second, MFPINet is designed to output 3D images at one time instead of  generating single focal plane image one-by-one like Deep-Z, which improves reconstruction efficiency. 

\section{Method}
In this part, we elaborate the network architecture, optimization functions and evaluation metrics of the proposed method. 
\subsection{Network Architecture}
\begin{figure*}[hb]
	\centerline{\includegraphics[scale=0.65]{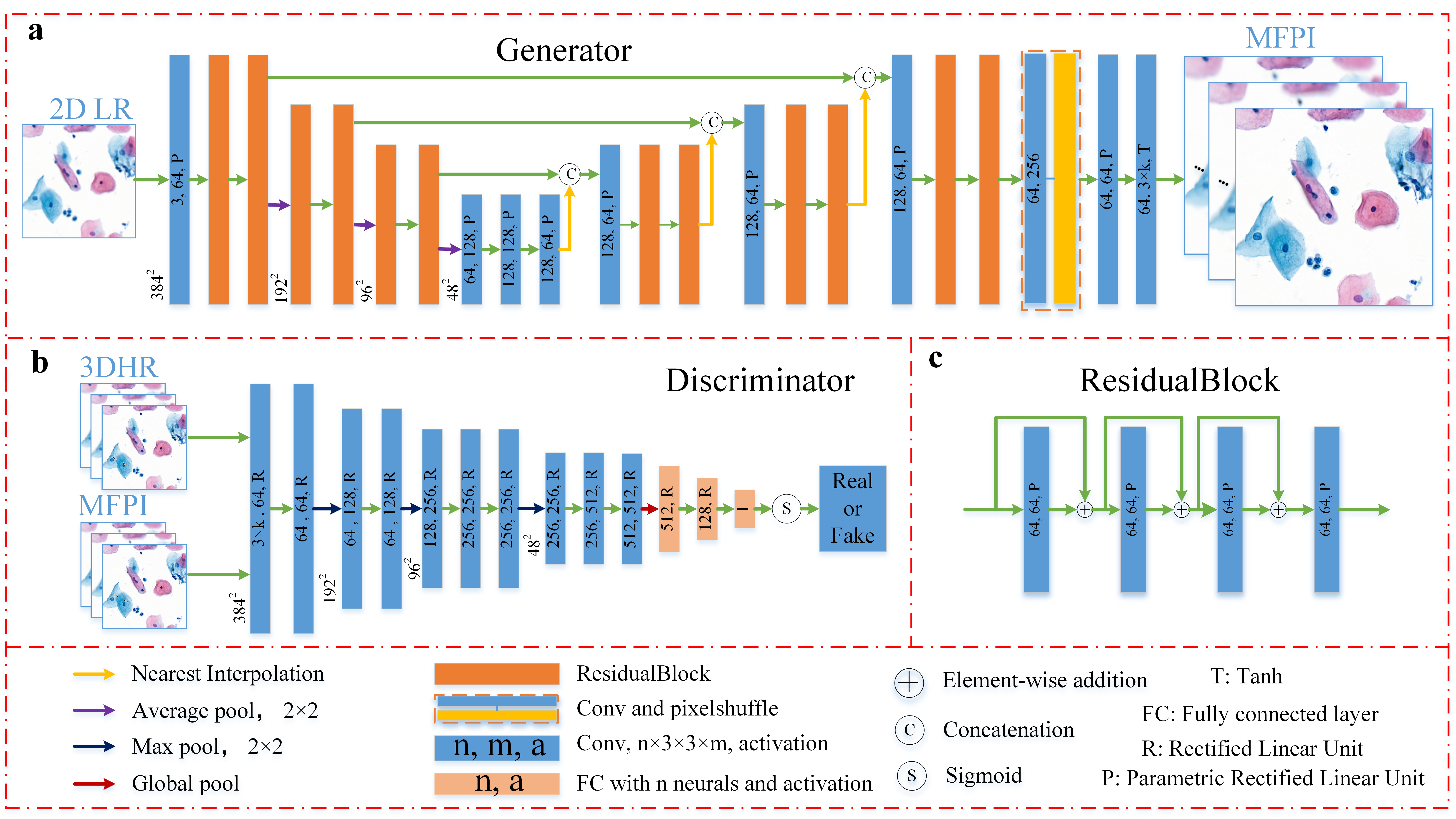}}
	\caption{The network architecture of MFPINet (example for $384\times 384\times 3$ sRGB image). The green arrow in the figure represents the flow of data, and the meaning of all other boxes and arrows is at the bottom of the figure. $\textbf{a}$. The architecture of the generator, which receives 2D wide-field images and outputs high-resolution MFP images. $\textbf{b}$. The architecture of the discriminator, which takes different MFP images as inputs and outputs corresponding categories of different images . $\textbf{c}$.  The basic module of the generator.}
	\label{figure1}
\end{figure*}
The proposed MFPINet is a GAN framework model, which consists of two parts: generator ($G$) and discriminator ($D$). Fig. \ref{figure1} shows the architecture of $G$ and $D$.

The generator consists of an encoder and a decoder, which is inspired by U-Net \cite{RN38} and ResNet \cite{RN41}.  The encoder is composed of a preprocessing module, three residual encoding modules (REM) and a stacked convolutional layers (SCL), which are expressed as the following equation:
\begin{equation}
x_k = SCL(REM(REM(REM(P(Conv_{64}(x_3))))))
\label{encoder:main}
\end{equation}
where $x_3$ stands for sRGB input image, $Conv_{k}(\cdot)$ stands for a convolutional layer with $k$ channels. $P(\cdot)$ stands for parametric rectified linear unit. $SCL(\cdot)$ stands for a stacked convolutional layers composed of three covolutional layers.  $REM$ represents the residual encoding module. $SCL$ and $REM$ can be expressed as equation \ref{encoder:SCL} and \ref{encoder:REM}:
\begin{equation}
SCL = Conv_{64}(Conv_{128}(Conv_{128}(x_k)))
\label{encoder:SCL}
\end{equation}
\begin{equation}
REM =AP(RB(RB(x_k)))
\label{encoder:REM}
\end{equation}
where $x_k$ stands for a feature map with k channels. $AP(\cdot)$ stands for 2D average pooling, which is used to halve the width and height of the feature map. $RB(\cdot)$ represents a residual block composed of four convolutional layers. Detailed topological connections can be found in Fig. \ref{figure1}. 

The encoder outputs feature maps of four different scales, which are $f_{1}$, $f_{2}$, $f_{3}$, $f_{4}$ from shallow to deep. Then decoder takes all output feature maps of the encoder as input for reconstruction. Specifically, the decoder consists of three residual decoding modules (RDM) and one super-resolution reconstruction (SRR) module, which are expressed as the following equation:
\begin{equation}
\begin{split}
MFPI = SRR(RDM(RDM(RDM( \\
f_{4}, f_{3}), f_{2}), f_{1}))
\end{split}
\label{decoder:main}
\end{equation}
where $RDM(\cdot)$ represents the residual decoder module, which is used to integrate feature information of different scales. $SRR(\cdot)$ represents the super-resolution reconstruction module, which is used to reconstruct HR MFP images from multi-scale feature maps fused by $RDM(\cdot)$. $SRR$ and $RDM$ are expressed as equation \ref{decoder:RDM} and \ref{decoder:SSR}:
\begin{equation}
RDM = RB(RB(CAT(NI(f_{k+1})), f_{k}))
\label{decoder:RDM}
\end{equation}
\begin{equation}
SSR = Tanh(Conv_{3\times n}(PS(Conv_{256}(f_{fused}))))
\label{decoder:SSR}
\end{equation}
where $f_k$ stands for the $k^{th}$ feature map output by the encoder. $NI(\cdot)$ stands for nearest interpolation at width and height dimension of feature maps. $CAT(\cdot)$ stands for the concatenation of tensors along channel dimension. $f_{fused}$ represents feature maps fused by the three $RDM$ modules. $PS(\cdot)$ represents pixel shuffle \cite{RN44} layer, which is used to double width and height dimension of feature maps. $Tanh(\cdot)$ represents tanh activation. $n$ represents the number of different focal plane 2D images in the output MFP images.

The above-mentioned encoder and decoder constitute the proposed generator. Although it borrows the U-shaped structure of U-Net, it is quite different from the original and common modified U-Net. First, the common BN module is discarded because BN is reported to cause artifacts \cite{wang2018esrgan}. Second, residual blocks are introduced to alleviate gradient problem caused by deep networks. Third, according to the characteristics of super-resolution tasks,  only three down-sampling blocks are kept. Because super-resolution tasks are different from high-level tasks such as semantic segmentation, pixels in super-resolution tasks generally only have a greater relationship with their neighboring pixels. At the same time, because the model becomes smaller, the model can process larger images at a faster speed, thereby enabling the model to quickly generate MFP images. Fourth, the post-sampling strategy is adopted instead of interpolating the input image at the beginning, which greatly reduces the consumption of memory and computational complexity. At last, the proposed generator directly outputs 3D MFP images instead of outputting refocused images \cite{RN23} one by one, which greatly improves the efficiency of acquiring MFP images.

The discriminator of the proposed MFPINet is a simple classification network, and its architecture is shown in Fig. \ref{figure1}. In order for the discriminator to perceive the difference of all focal plane images at the same time, the designed discriminator takes 3D MFP images as input. In addition, to reduce consumption of memory and computation, 3D MFP images are converted into 2D images, and then use 2D convolution to process them. Specifically, the original 3D MFP image represented by a four-dimensional array $H\times W\times Z\times C$ is converted into a three-dimensional array $H\times W\times (Z\times C)$.

\subsection{Model Optimization}
The proposed generator is designed to learn a transformation mapping from LR 2D images to corresponding 3D HR MFP images. In the training phase, generator $G_{\theta_g}$ with parameter $\theta_g$ takes a 2D LR image $X$ as the input and output 3D HR MFP images $G_{\theta_g}(X)$. The discriminator $D_{\theta_d}$ with parameter $\theta_d$ takes real HR MFP images and forged MFP images by  $G_{\theta_d}$ respectively. Discriminator $D_{\theta_d}$ outputs a probability value for input MFP images, which represents the model's confidence. The loss $L_G$ of generator is defined as following equation:
\begin{equation}
\begin{split}
L_G = \frac{1}{N}\sum\limits_{n=1}^{N}{-log(D_{\theta_d}(G_{\theta_g}(X^{\{n\}})))}\times\alpha \\
+ |G_{\theta_{g}}(X^{\{n\}})-Y^{\{n\}}| 
\end{split}
\label{loss:LG}
\end{equation}
where $N$ represents the number of samples of mini-batch in each iteration. $X^{\{n\}}$  is the $n^{th}$ 2D input image.  $G_{\theta_g}{(X^{\{n\}})}$ is the HR MFP images obtained by the generator, and $Y^{\{n\}}$  is the real HR MFP images corresponding to $ X^{\{n\}}$. $\alpha$ is a hyper-parameter used to balance absolute error loss and adversarial loss, which is set to 0.01 in the experiments.

As for the discriminator loss $L_D$, it is defined as  the following equation:
\begin{equation}
\begin{split}
L_D = -\frac{1}{N}\sum\limits_{n=1}^{N}{log(D_{\theta_d}(Y^{\{n\}}))} \\
+ log(1- D_{\theta_d}(G_{\theta_g}(X^{\{n\}})))
\end{split}
\label{loss:LD}
\end{equation}
The meaning of the parameters in equation \ref{loss:LD} is similar to that in equation \ref{loss:LG}. In order to improve the training speed and optimization stability, Adam \cite{RN45} optimization algorithm is used to optimize both generator $G_{\theta_g}$ and discriminator $D_{\theta_d}$. The learning rate is set to 0.0001 and other parameters are set to the default value. More details can be found in \textbf{\itshape{Experiments}} section.

\subsection{Evaluation Metrics}
In order to evaluate the quality of images generated by MFPINet, four common metrics are used to compare the differences between the generated MFP images $I_g$ and the real mechanical scanning MFP images $I_{gt}$. In order to explore the reconstruction effect of different focal planes, the MFP images are not evaluated as a whole, but as individual focal plane.Specifically, the following four metrics are used.
\subsubsection{Mean Square Error}

Mean square error (MSE) is widely used to evaluate the error of two sets. For evaluation of MFP images, first split it into $N$ 2D images along the axial axis, and then $N$ corresponding real mechanical scanning 2D images and the $N$ reconstructed 2D images are evaluated separately. Specifically, MSE is defined as following equation:
\begin{equation}
MSE(I_{g}, I_{gt}) = \frac{1}{N}\sum\limits_{n=1}^{N}{\frac{1}{H\times W\times C}{||I_{g}^{(n)}-I_{gt}^{(n)}||_{2}^{2}}}
\label{evaluation:mse}
\end{equation}
where $H$ and $W$ are the height and width of images. $C$ is the channel number of images (e.g. 3 for sRGB image). $N$ is the total number of images. Regarding MSE, the smaller the value, the better the generated image.

\subsubsection{Peak Signal to Noise Ratio}
Peak signal to noise ratio (PSNR) is the most widely used objective method to evaluate image quality. Similar to MSE, PSNR is defined as following equation:
\begin{equation}
PSNR(I_g, I_{gt})=10\times log_{10}{(\frac{(2^n-1)^2}{MSE(I_g, I_{gt})})}
\label{evaluation:psnr}
\end{equation}
where $n$ is the bits number of an image (e.g. 8 for uint8). Compared with MSE, the larger the PSNR value, the better the generated image.
\subsubsection{Mean Absolute Error}
Mean absolute error (MAE) uses L1 distance, which is less sensitive to outliers than L2 distance of MSE. It is defined as follows:
\begin{equation}
MAE(I_{g}, I_{gt}) = \frac{1}{N}\sum\limits_{n=1}^{N}{\frac{1}{H\times W\times C}{||I_{g}^{(n)}-I_{gt}^{(n)}||_{1}}}
\label{evaluation:mae}
\end{equation}
The meaning of parameters in equation \ref{evaluation:mae} is the same as that in equation \ref{evaluation:mse}. Similar to MSE, the smaller the MAE value, the better the generated image.
\subsubsection{Structure Similarity Index}
Structural similarity index (SSIM) is a metric to measure the similarity of two images. It considers more attributes of images, such as brightness, contrast and structure similarity. So it is more consistent with human vision than PSNR or MSE. SSIM is defined as following equation:
\begin{equation}
SSIM(I_g, I_{gt}) = \frac{(2\mu_{g}\mu_{gt}+c_1)(2\sigma_{g,gt}+c_2)}{(\mu_{g}^{2}+\mu_{gt}^{2}+c_1)(\sigma_{g}^{2}+\sigma_{gt}^{2}+c_2)} 
\end{equation}
where $\mu_g$ and $\mu_{gt}$ are average brightness of the generated image $I_g$ and mechanical scanning image $I_{gt}$ respectively. $\sigma_g$ and $\sigma_{gt}$ are standard deviation of $I_g$ and $I_{gt}$ respectively. $\sigma_{g,gt}$ is covariance of two images. $c_1$ and $c_2$ are constants used to stabilize the division process.

\section{Experiments}
To verify the effectiveness of MFPINet, we evaluate the quality of  images generated by MFPINet in cytology images. As there is no method that fully matches our work at present, we compare our method with Deep-Z \cite{RN23} in reconstructing 3D images. At the same time, to evaluate the super-resolution effect of our model, the classical SRGAN \cite{ledig2017photo} is used as a baseline.
\subsection{Cytology Images}
\subsubsection{Data Collection}
In our experiments, 5 liquid-based cytology slides are used, which come from the department of clinical laboratory, Tongji Hospital, Huazhong University of Science and Technology. For each glass slide, 11 digital whole image slides (WSI) with a step of 2.7 $\mu m$ are scanned under 20$\times$/0.75-NA objective lens using 3D Histech scanner. The resolution of WSIs is 0.243 $\mu m/pixel$ with a size of about 100K × 100K pixels. 

\subsubsection{Data Preparation}
The 5 slides are randomly divided into training set (3 slides) and test set (2 slides). For each slide, 5000 patches of 768$\times$768 size with cell foreground are cropped out as ground truth (GT). Then bicubic interpolation are used to resize the patch size from 768$\times$768 into 384$\times$384. In order to further reduce the quality of input data, Gaussian blur with a window of 5$\times$5 and a standard deviation of 3 is used to degrade the input LR images. Finally, the input LR images and HR GT images are normalized to -1$\sim$1. In experiments, the middle LR layer image (the 6th of ll layers, defined as 0 $\mu$m) is taken as the input and all 11 HR layers are taken as the GT. 

\subsection{Training Details}
All experiments are completed in PyTorch 1.4 \cite{RN46} and Ubuntu 18.04 equipped with two NVIDIA RTX 2080ti GPUs. For SRGAN and Deep-Z, we retrained these methods using the parameters provided in original papers on our dataset.
\subsubsection{MFPINet}
When training MFPINet, Adam optimizer is used with an initial learning rate 0.0001. The model is trained 20 epochs on the training set and the learning rate is halved every two epochs. 
\subsubsection{SRGAN}
When training SRGAN, the GT and input are the middle layer HR image (0 $\mu m$) and corresponding degraded LR image respectively. The degraded method is consistent with our model.
\subsubsection{Deep-Z}
As for Deep-Z, the input is a $384\times384$ HR image and a digital propagation matrix (DPM, used to indicate the focal depth). GT is a HR image of specific  focal plane corresponded to DPM.
\subsection{Experiment Results}
In this part, we show the overall effect of our model, as well as the individual effect of horizontal super resolution and axial refocusing.

\subsubsection{Multi-focal Plane Imaging Using MFPINet}
{
\begin{figure*}[hb]
	\centerline{\includegraphics[scale=0.165]{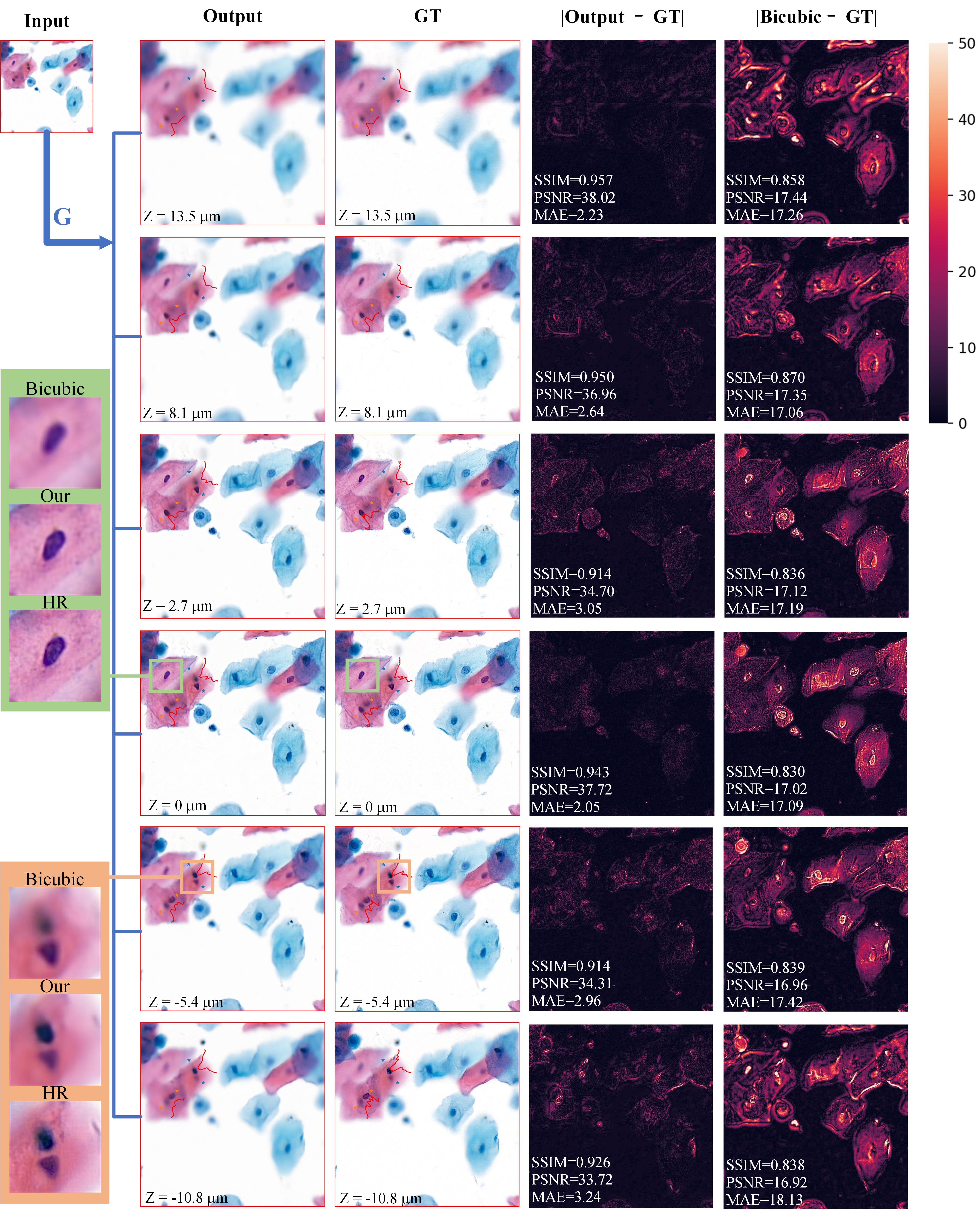}}
	\caption{The process and results of multi-focal plane imaging using MFPINet. Columns from left to right are horizontal super resolution of different positions, reconstructed HR MFP images by MFPINet, mechanical scanning HR MFP images \textit{GT}, reconstruction error maps between MFPINet’s output and \textit{GT}, reconstruction error maps between the interpolated input and \textit{GT}. The red curve in the second and third columns represents the gray intensity distribution of pixels between two points.}
	\label{figure2}
\end{figure*}

Fig. \ref{figure2} shows the reconstructed HR MFP images from a single 2D LR image. The focal depth range of the reconstructed MFP images is -13.5 $\mu m$ to 13.5 $\mu m$ with an interval of 2.7 $\mu m$. The \textit{Output} and \textit{GT} columns in Fig. \ref{figure2} represent HR MFP images of MFPINet reconstruction and mechanical scanning separately. The right two columns in Fig. \ref{figure2} are reconstruction error maps. $| \textit{Output}-\textit{GT}|$ represents the MAE between MFPINet’s output and GT. $|\textit{Bicubic}-\textit{GT}|$ represents the MAE between the interpolated input and GT. SSIM, PSNR and MAE in the error maps are based on the current images. From the error maps and corresponding metrics, it can be seen that MFPINet can effectively reconstruct HR MFP images. In addition, the red curves in \textit{Output} and \textit{GT} columns represent the gray intensity distribution of pixels between two points. The results indicate that MFPINet can reconstruct well local details. On the left side of Fig. \ref{figure2}, some small patches were taken from different positions. The interpolated LR images \textit{Bicubic}, \textit{Output} and \textit{GT} patches corresponding to the same position were spliced together. The results reflect the super-resolution effect of MFPINet. Notably, the error maps are obtained on corresponding gray images and the metrics were computed on sRGB images.

We provided the reconstructed images of all focal planes of Fig. \ref{figure2} in \textit{Supplementary A}. Besides the reconstruction error map, we also analyzed the reconstruction quality from frequency spectrum. In addition, we further analyzed the influence of input image resolution on axial refocusing reconstruction. The experimental results (see \textit{Supplementary B}) show that the reconstruction effect of MFPINet with HR input is better than that of MFPINet with LR input. This indicates that LR input increases the difficulty of axial refocusing reconstruction. 
}

\subsubsection{Super Resolution}
{
MFPINet combines the reconstruction of MFP images and horizontal super resolution. In order to verify the individual effect of super resolution, we compared SRGAN and our method on the test set. Table \ref{table:sr} shows the statistical values on SSIM, PSNR, MSE and MAE metrics. The results indicate that MFPINet performs slightly better than SRGAN on the four metrics. Fig. \ref{figure3}  shows the images generated by the two models. The left four columns are the original sRGB images, and the right three columns correspond to reconstruction error maps with SSIM, PSNR and MAE metrics. From the error maps, it can be seen that our method has smaller error in cytoplasm while SRGAN has smaller error in nucleus. Overall, our method is consistent with SRGAN. The results demonstrate that MFPINet is effective for horizontal super resolution while generating MFP images. Notably, SRGAN was trained and tested on the middle layer (0 $\mu m$) images.

\begin{table}{}
\centering
\begin{threeparttable}
	\caption{Comparison of SRGAN and MFPINet}
	\label{table:sr}
	\begin{tabular}{ccccc}
		\toprule
		Method & SSIM & PSNR & MSE & MAE\cr
		\midrule
		SRGAN & 0.952 & 38.05 & 10.67 & 2.16 \cr
		MFPINet & 0.954 & 38.63 & 9.65 & 1.79 \cr
		\bottomrule
	\end{tabular}
\end{threeparttable}
\end{table}

\begin{figure*}[h]

	\centerline{\includegraphics[scale=0.21]{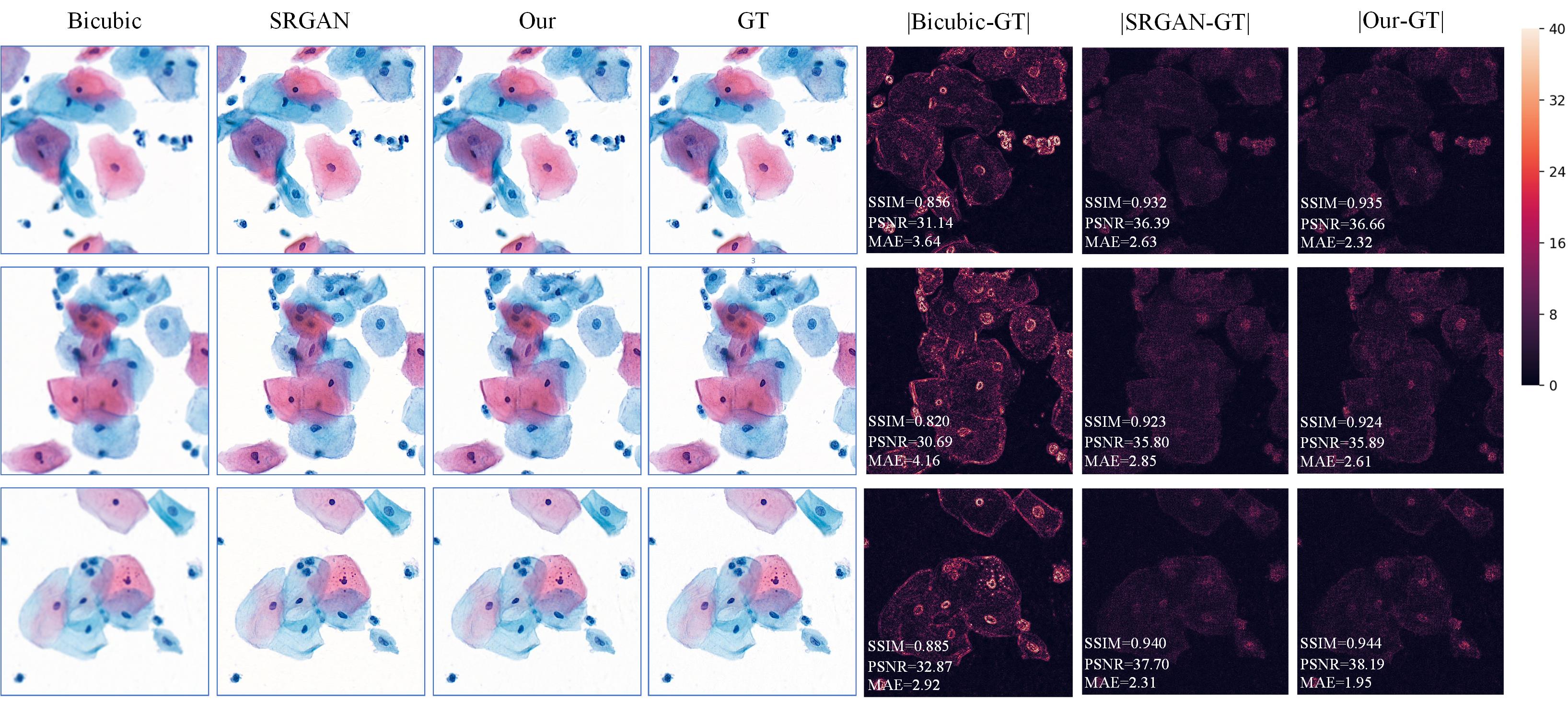}}
	\centering
	\caption{HR images generated by bicubic interpolation, SRGAN and our method. The images of our method shown in the figure are taken from 0 $\mu m$ layer of reconstructed MFP images by MFPINet. The meaning of other elements in the figure is similar to Fig. \ref{figure2}}
	\label{figure3}
\end{figure*}
}

 \subsubsection{Multi-focal Plane Reconstruction}
 {
 In this section, we compared Deep-Z and our method on the test set to demonstrate the refocusing ability of MFPINet. Table \ref{table:refocus} shows the statistical values of four metrics.
 \begin{table}{}
  \centering  
\fontsize{6.5}{8}\selectfont  
\begin{threeparttable}  
	\caption{Comparison of Deep-Z and MFPINet}  
	\label{table:refocus}
	\setlength{\tabcolsep}{4pt}
	\begin{tabular}{ccccccccc}  
		\toprule  
		\multirow{2}{*}{Focal Depth}&  
		\multicolumn{4}{c}{Deep-Z}&\multicolumn{4}{c}{MFPINet}\cr  
		\cmidrule(lr){2-5} \cmidrule(lr){6-9}  
		&PSNR&SSIM&MSE&MAE&PSNR&SSIM&MSE&MAE\cr  
		\midrule  
		-13.5$\mu$m&37.01&0.942&17.61&2.42&38.20&0.957&12.30&2.02\cr  
		-10.8$\mu$m&36.92&0.942&18.32&2.40&37.91&0.955&13.33&2.04\cr
		-8.1$\mu$m&36.71&0.938&19.25&2.41&37.46&0.950&14.67&2.11\cr
		-5.4$\mu$m&36.43&0.933&19.86&2.45&36.98&0.944&15.68&2.19\cr
		-2.7$\mu$m&36.75&0.935&17.13&2.36&36.78&0.940&15.43&2.22\cr
		0.0$\mu$m&46.50&0.993&1.64&0.73&38.64&0.954&9.65&1.79\cr
		2.7$\mu$m&35.78&0.922&20.53&2.60&36.00&0.928&17.90&2.39\cr
		5.4$\mu$m&36.67&0.932&17.13&2.42&37.63&0.944&12.80&2.15\cr
		8.1$\mu$m&38.22&0.953&11.65&2.11&39.56&0.963&7.82&1.84\cr
		10.8$\mu$m&38.68&0.957&10.36&2.03&40.24&0.967&6.56&1.73\cr
		13.5$\mu$m&38.66&0.958&10.45&2.04&40.32&0.968&6.42&1.72\cr
		\bottomrule  
	\end{tabular}  

\end{threeparttable}  
\end{table}  
As can been seen from the table, MFPINet slightly exceeds Deep-Z on all focal planes except the layer 0 $\mu m$. This phenomenon is caused by different inputs. The input of Deep-Z is the layer 0 $\mu m$ HR image, while the input of MFPINet is the layer 0 $\mu m$ LR image. For the focal plane of 0 $\mu m$, Deep-Z degenerates into identity mapping, while our method degenerates into super resolution. So the effect of Deep-Z is much higher than our method in this focal plane. In order to further explore the difference between the two methods, we visualized their generated images in Fig. \ref{figure4}.
\begin{figure*}[h]
	\centerline{\includegraphics[scale=0.27]{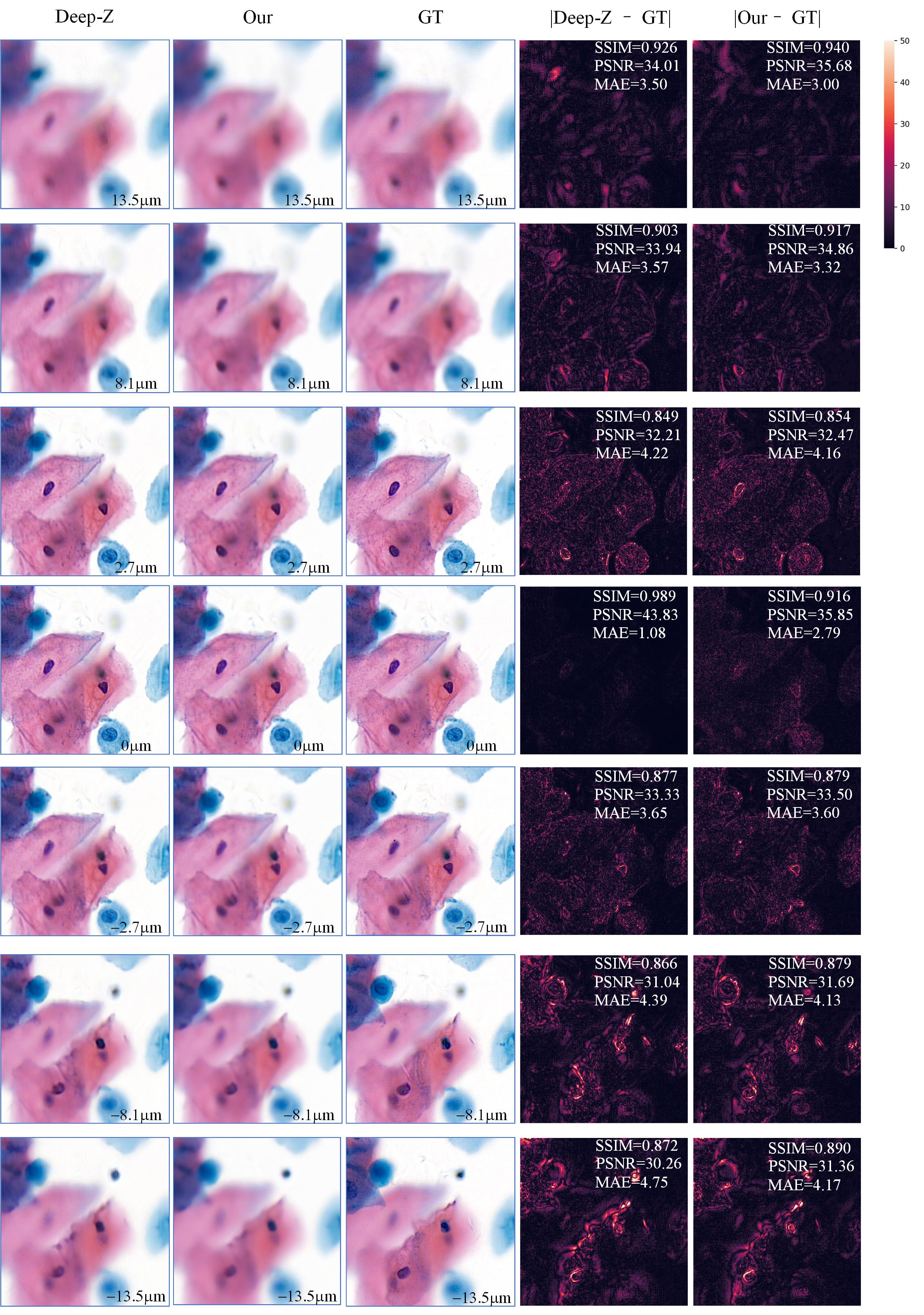}}
	\centering
	\caption{MFP images generated by Deep-Z and our method. The meaning of other elements in the figure is similar to Fig. \ref{figure2}}
	\label{figure4}
\end{figure*}
The left three columns in Fig. \ref{figure4} represent the original sRGB images and the right two columns represent reconstruction error maps. It can be seen that our method perform better than Deep-Z on all focal planes except the layer 0 $\mu m$. The error maps also show that the images reconstructed by the two methods have their own characteristics. Deep-Z's reconstructed nuclei near the layer 0 $\mu m$ are slightly better. Because the input image of Deep-Z has not been degraded and adjacent focal planes can learn more details from the HR input. However, as the focal plane becomes farther and farther, Deep-Z gradually loses the advantage of HR input, and the quality of generated images decreases. On the contrary, MFPINet is better than Deep-Z on non-middle focal planes. Notably, the images on focal planes more than 10 $\mu m$ from the middle generally are overall blurred, thus showing higher metrics. In summary, although the input of our model is a LR image, the quality of reconstructed MFP images is consistent with that of Deep-Z. This results indicates that it is feasible to reconstruct high-quality MFP images using a single LR wide-field image.
}

\subsubsection{Computational Efficiency}
The purpose of the proposed MFPINet is to speed up 3D imaging. In order to verify the actual performance of MFPINet, we analyzed the complexity of MFPINet and tested the actual running time. Specifically, we tested the running time for obtaining a 768$\times$768$\times$11$\times$3 size MFP image on a single RTX 2080ti GPU, and analyzed parameters and giga floating-point operations per second (GFLOPS) of MFPINet. Table \ref{table:speed} shows the results. \textit{Time} column is the average time of one thousand calculations. It can be seen that MFPINet is approximately 24 times faster than Deep-Z when reconstructing 11-layer MFP images of the same size. At the same time, the parameters of MFPINet is only 1/8 of Deep-Z. In summary, the designed MFPINet is efficient and has a potential to be applied in actual scenarios.

\begin{table}{}
	\centering
	\begin{threeparttable}
		\caption{Computational efficiency of Deep-Z and MFPINet}
		\label{table:speed}
		\begin{tabular}{cccc}
			\toprule 
			Method & GFLOPS & Params (M)& Time (ms)\cr
			\midrule
			Deep-Z (Single layer) & 762.4 & 19.41 & 61.3 \cr
			Deep-Z (MFP) &8386.4 & 19.41 & 673.8 \cr
			MFPINet & 264.2 & 2.50 & 27.8 \cr
			\bottomrule
		\end{tabular}
	\end{threeparttable}
\end{table}

\section{Conclusion}
The acquisition speed of HR 3D images is limited by axial scanning. To address the challenge, we propose a generative adversarial network named MFPINet to reconstruct HR MFP images from a single 2D wide filed microscopic image. The generator and discriminator are designed to refocus all focal planes at one time and post-sampling strategy is adopted to improve reconstruction efficiency. We conduct extensive experiments on cytology microscopy images and demonstrate that MFPINet performs well on both axial refocusing and horizontal super resolution. Experiments further show MFPINet is approximately 24 times faster than current refocusing methods for reconstructing the same volume images. 


We demonstrate that MFPINet can reconstruct a 3D image with a depth of 29.7 $\mu m$ from a single 2D image, but the experimental results show that the reconstruction effect far away from the center layer will be attenuated. Therefore, how to reconstruct deeper 3D images with high quality is still a challenge. For example, the density of objects in input images may influence the effect and depth of 3D reconstruction. We will further explore the problem in the future.

\section*{Acknowledgment}
We thank the Optical Bioimaging Core Facility of Wuhan National Laboratory for Optoelectronics, Huazhong University of Science and Technology for the support in data acquisition.

\bibliographystyle{IEEEtran}
\bibliography{IEEEabrv,reference.bib}

\end{document}